\documentclass[prb,aps,showpacs,twocolumn,floatfix]{revtex4}
\usepackage{amsmath,amsfonts,amssymb,graphicx,color,bm}
\usepackage[english]{babel}
\newcommand{\bi}{{\bm i}}
\newcommand{\bj}{{\bm j}}
\newcommand{\bk}{{\bm k}}
\newcommand{\bl}{{\bm l}}
\newcommand{\bS}{{\bm S}}
\newcommand{\xh}{\hat{x}}
\newcommand{\yh}{\hat{y}}

\begin{document}
\title{Phase diagram of the Heisenberg antiferromagnet with four-spin
       interactions}
\author{L. Isaev$^1$}
\author{G. Ortiz$^1$}
\author{J. Dukelsky$^2$}
\affiliation{$^1$Department of Physics, Indiana University, Bloomington IN
                 47405, USA \\
             $^2$Instituto de Estructura de la Materia - CSIC, Serrano 123,
                 28006 Madrid, Spain}
\begin{abstract}
 We study the quantum phase diagram of the Heisenberg planar antiferromagnet
 with a subset of four-spin ring exchange interactions, using the recently
 proposed heirarchical mean-field approach. By identifying relevant degrees of
 freedom, we are able to use a single variational anzatz to map the entire
 phase diagram of the model and uncover the nature of its various phases. It is
 shown that there exists a transition between a N\'eel state and a quantum
 paramagnetic phase, characterized by broken translational invariance. The
 non-magnetic phase preserves the lattice rotational symmetry, and has a
 correlated plaquette nature. Our results also suggest that this phase
 transition can be properly described within the Landau paradigm.
\end{abstract}
\maketitle
\section{Introduction}

Quantum phases of matter and their transitions are of fundamental concern to
modern condensed matter physics \cite{Sachdev_1999}. Such interest is motivated
not only by potential technical applications, but also on purely scientific
grounds. Research in this field may lead to a deeper understanding of the
fundamental working principles behind Nature's behavior, and often original new
physical theories emerge. One of them, recently proposed in Ref.
\onlinecite{Sachdev_2004}, predicts the existence of a class of systems whose
critical behavior lies outside the scope of the Landau theory of phase
transitions \cite{Landau_1999}. Critical points in these systems are
characterized by the deconfinement of fractionalized excitations,
parameterizing the original degrees of freedom, which occurs right at the
transition. It was observed that this scenario can, in principle, be realized
in spin systems, which exhibit a second-order phase transition point
characterized by the simultaneous breakdown of a continuous (e.g. spin $SU(2)$)
and a discrete (e.g. lattice) symmetries, in such a way that symmetry groups on
opposite sides of the transition are not group-subgroup related. Such critical
points cannot be described in the framework of Landau's theory.

According to Ref. \onlinecite{Sachdev_2004}, there should exist a substantial
number of spin systems, which exhibit deconfined critical points. For instance,
frustrated two-dimensional (2D) antiferromagnets (AFs), like the $J_1$-$J_2$
model, are believed to fall into this category. However, there seems to be no
experimental or theoretical proof of this claim. Another class of models
believed to display such a behavior includes non-frustrated AFs with multi-spin
exchange interactions. One such model was studied by Sandvik
\cite{Sandvik_2007} and other authors \cite{Melko_2008_prl,Melko_2008_prb} and,
although seemingly artificial, it provides a playground for testing new
theories. Their Quantum Monte Carlo (QMC) simulations claimed numerical
evidence for the deconfined quantum criticality scenario. 

The model studied in Ref. \onlinecite{Sandvik_2007} is a Heisenberg AF with a
subset of four-spin ring exchange interactions, defined on a square lattice
(named the $J$-$Q$ model)
\begin{displaymath}
 H=J\sum_{\langle\bi\bj\rangle}\bS_\bi\bS_\bj-Q\sum_{\langle\bi\bj\bk\bl
 \rangle}\biggl(\bS_\bi\bS_\bj-\frac{1}{4}\biggr)\biggl(\bS_\bk\bS_\bl-
 \frac{1}{4}\biggr),
\end{displaymath}
where $Q\geqslant0$, $\bi,\bj,\ldots$ denote sites in a 2D square lattice and
$\bS_\bi$ are spin-$1/2$ operators. The first summation extends over bonds
(nearest neighbor sites). The second term contains two sums over plaquettes
(sites of the dual lattice): first, $(\bi\bj)$ and $(\bk\bl)$ denote parallel
horizontal links of the plaquette, and then $(\bi\bk)$ and $(\bj\bl)$
correspond to parallel vertical bonds. It was concluded \cite{Sandvik_2007}
that there exists a critical point at $Q_c/J\sim25$ separating the
antiferromagnetic phase from a valence-bond solid (VBS) state, whose nature is,
strictly speaking, unclear \cite{Sandvik_2007} but the calculations suggested a
columnar (dimer) order in this paramagnetic region.

In the present paper we study the phase diagram of the $J$-$Q$
model, using a recently proposed hierarchical mean-field (HMF) technique
\cite{Isaev_2009,Ortiz_2003}. The main idea of the method revolves around the
concept of a {\it relevant} degree of freedom (a ``quark'') -- spin
cluster in this particular case -- which can be used to build up the
system. The initial Hamiltonian is then rewritten in terms of these
coarse-grained variables and a mean-field approximation is applied to
determine properties of the system. Thus, the (generally) exponentially hard
problem of determining the ground state of the system is reduced to a {\it
polynomially} complex one. At the same time, essential quantum correlations,
which drive the physics of the problem, are captured by the local
representation. In other words, provided the quark is chosen properly, even a
simple {\it single} mean-field approximation, performed on these degrees of
freedom, will yield the correct phase diagram. Moreover, our HMF ansatz
provides an {\it educated nodal surface} that can, in principle, be used in
conjunction with fixed-node (or constrained path \cite{Carlson_1999}) QMC
approaches to further improve correlations, and thus energy estimates, in those
cases when there is a sign (phase) problem.

It is important to emphasize the simplicity of our method. In this work we
concentrate on symmetries of the various phases, exhibited by the $J$-$Q$
model. By using a more sophisticated variational ansatz (e.g. a Jastrow--type
correlated wavefunction), one can also improve numerical values of the
observable quantities and phase transition points, but the physical picture
will remain intact. Nevertheless, the HMF method was quite accurate to yield
the quantitatively correct phase diagram \cite{Isaev_2009} of the $J_1$-$J_2$
model, whose behavior is driven by the interplay of two gapless phases: the
N\'eel and columnar AF states. In the $J$-$Q$ model the large-$Q$ phase is
gapped. Due to its real-space nature, the HMF method should be appropriate for
this model. Indeed, our recent studies \cite{SS_2009} of another gapped system
-- the Heisenberg model on the Shastry-Sutherland lattice -- support this
assumption.

\begin{figure}[!t]
 \begin{center}
  \includegraphics[width=\columnwidth]{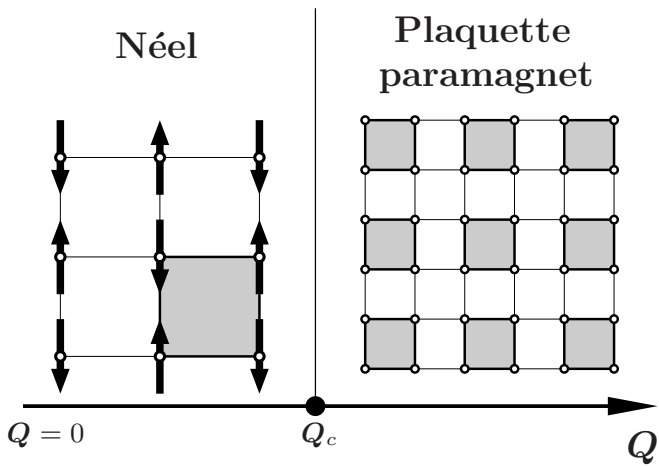}
 \end{center}
 \caption{Schematic phase diagram of the $J$-$Q$ model, obtained in the
	  present paper. The translationally invariant AF phase with broken
	  $SU(2)$ symmetry and the singlet paramagnetic state are separated by
          a quantum phase transition at $Q=Q_c$.
          Results of our calculations indicate that this transition is
          continuous, but we can not rigorously discard the possibility
          that it becomes weakly first order in the thermodynamic limit. The
          non-magnetic phase breaks the lattice translational
          symmetry and is a {\it plaquette paramagnet}.}
 \label{fig_phase_diag}
\end{figure}

Our findings are summarized in Fig. \ref{fig_phase_diag}. There indeed exists a
non-magnetic phase, which we found to be of a {\it correlated plaquette} type,
and not of a dimer character, separated from the N\'eel state by a second-order
phase transition. However, the numerical value of the critical point,
$Q_c/J\sim2$, which we obtained, is quite different from that of Ref.
\onlinecite{Sandvik_2007}. Our results are consistent with data obtained from
exact diagonalization of finite spin clusters which, given the fact that the
system is gapped in the paramagnetic phase for $Q\geqslant Q_c$, should be
reliable in this region. Although we found the phase transition to be of the
Landau second-order type, due to the real-space nature of the method (which
explicitly breaks the lattice translational invariance), we cannot rigorously
rule out the possibility that this phase transition becomes weakly first order
as one scales the degree of freedom towards the thermodynamic limit.

In the next section we set up the formalism. The results are presented in Sec.
\ref{section3}, followed by a discussion in Sec. \ref{section4}.

\section{Coarse graining and HMF approximation}
\label{section2}

For our purposes it is convenient to separate the two and four spin terms in
the $J$-$Q$ Hamiltonian:
\begin{align}
 H=&-\frac{2NQ}{16}+\biggl(J+\frac{Q}{2}\biggr)\sum_{\langle\bi\bj\rangle}
 \bS_\bi\bS_\bj- \nonumber \\
 &-Q\sum_{\langle\bi\bj\bk \bl\rangle}(\bS_\bi\bS_\bj)(\bS_\bk\bS_\bl).
 \label{initial_hamiltonian}
\end{align}

A satisfactory coarse graining procedure should partition the lattice into spin
clusters (quarks), containing $N_q$ sites, that explicitly preserve symmetries
of the Hamiltonian. In particular, the $J$-$Q$ Hamiltonian is explicitly
spin--$SU(2)$ invariant. Moreover, it is invariant under transformations from
the lattice rotational group $C_4$. Therefore, we will consider only symmetry
preserving degrees of freedom: ({\it i}) plaquettes ($2\times2$ spin clusters)
and ({\it ii}) $4\times4$ spin clusters. Each cluster state will be associated
with a hard-core bosonic operator $\gamma$. These operators are Schwinger
bosons of $SU\bigl(2^{N_q}\bigr)$ and must obey the local constraint:
$\sum_{a}\gamma^\dag_{ia}\gamma_{ia}=1$. They define the hierarchical language
\cite{Ortiz_2003} for our problem.

From the form of Eq. \eqref{initial_hamiltonian} it is clear \cite{Isaev_2009}
that the (equivalent and exact) bosonic Hamiltonian will contain not only
two-body scattering processes, but also four-boson interactions. Therefore, we
can write down symbolically:
\begin{align}
 H=&\sum_{i}\bigl(H_\Box\bigr)_{a^\prime a}\gamma^\dag_{ia^\prime}\gamma_{ia}+
 \nonumber \\
 &+\sum_{\langle ij\rangle}\bigl(H^2_{\rm int}\bigr)^{a^\prime_1a^\prime_2}_
 {a_1a_2}\gamma^\dag_{ia_1^\prime}\gamma^\dag_{ja_2^\prime}\gamma_{ia_1}
 \gamma_{ja_2}+ \nonumber \\
 +&\sum_{\langle i_1i_2i_3i_4\rangle_h}\bigl(H^{4h}_{\rm int}\bigr)^{a^\prime_1
 a^\prime_2;a^\prime_3a^\prime_4}_{a_1a_2;a_3a_4}\prod_{\mu=1}^4
 \gamma^\dag_{i_\mu a_\mu^\prime}\gamma_{i_\mu a_\mu}+ \nonumber \\
 +&\sum_{\langle i_1i_2i_3i_4\rangle_v}\bigl(H^{4v}_{\rm int}\bigr)^{a^\prime_1
 a^\prime_3;a^\prime_2a^\prime_4}_{a_1a_3;a_2a_4}\prod_{\mu=1}^4
 \gamma^\dag_{i_\mu a_\mu^\prime}\gamma_{i_\mu a_\mu},
 \label{initial_bosonic_hamiltonian}
\end{align}
where $a,\ldots$ label states in the Hilbert space of a quark, $i,j,\ldots$
denote sites in the coarse grained lattice, and summations are assumed over all
repeated indices. The term with $H^2_{\rm int}$ encodes two-body interactions,
while the last two lines describe the correlated four-boson scattering. The
superscript $h$ indicates that $i_1i_2$ and $i_3i_4$ are horizontal links of a
plaquette, and similarly $v$ denotes the case when $i_1i_3$ and $i_2i_4$ are
vertical links of the same plaquette.

We will investigate the phase diagram of the $J$-$Q$ model using the HMF
approximation whose variational state assumes that the hard-core bosons form an
insulating state. Further, we introduce a new set of bosonic operators, related
to the old ones by a real site-independent canonical transformation:
\begin{displaymath}
 \gamma_{ia}=R_a^n\Gamma_{in};\,\,\,R_a^nR_b^n=\delta_{ab},\,\,\,
 R_a^nR_a^m=\delta_{nm}
\end{displaymath}
and write the variational ground state in the form:
\begin{equation}\label{HF_wavefunction}
 |\psi_0\rangle=\prod_i\Gamma^\dag_{i0}|0\rangle,
\end{equation}
where $n=0$ denotes the lowest energy single-particle mode (we shall also
denote: $R_a^0\equiv R_a$), and $|0\rangle$ represents the vacuum. It is
important to emphasize that although the coarse graining procedure preserves
the symmetries of the Hamiltonian, some of them can be spontaneously broken at
the mean-field level as a result of self-consistency. In particular, the
columnar dimer state is contained in the wavefunction \eqref{HF_wavefunction}
although, as we will see below, it never appears as a stable solution.

We have explicitly separated the four-boson interaction in the Hamiltonian
\eqref{initial_bosonic_hamiltonian} into horizontal and vertical link
contributions. This distinction is important because these two terms must be
properly symmetrized to fulfill bosonic statistics. In particular, the term
$H^{4h}_{\rm int}$ has to be symmetrized only with respect to indices in the
same group, and groups as a whole (groups are separated by semicolons), i.e.
one needs to take into account only the following permutations:
$(1\leftrightarrow2)$, $(3\leftrightarrow4)$ and simultaneously
($1\leftrightarrow3$, $2\leftrightarrow4$). Analogously, in the term
$H^{4v}_{\rm int}$ only the permutations $(1\leftrightarrow3)$,
$(2\leftrightarrow4)$ and ($1\leftrightarrow2$, $3\leftrightarrow4$) should be
accounted for.

The problem then reduces to minimization of the energy functional:
\begin{align}
 \frac{N_q E_0[R]}{N}=&(H_\Box)_{a^\prime a}R_{a^\prime }R_a+ \nonumber \\
 &+\bigl(H^2_{\rm int}\bigr)^{a^\prime_1a^\prime_2}_{a_1a_2}\prod_{\nu=1}^2
 R_{a_\nu^\prime}R_{a_\nu}+ \label{energy_functional} \\
 &+\bigl(H^{4h}_{\rm int}+H^{4v}_{\rm int}\bigr)^{a^\prime_1a^\prime_2;
 a^\prime_3a^\prime_4}_{a_1a_2;a_3a_4}\prod_{\nu=1}^4R_{a_\nu^\prime}R_{a_\nu}
 \nonumber
\end{align}
under the constraint $R_aR_a=1$, which leads to the self-consistent eigenvalue
equation:
\begin{subequations}\label{mean_field_eq}
 \begin{equation}\label{hf_equation}
  \bigl(H_{HF}\bigr)_{ab}R_b=\mu R_a
 \end{equation}
 with the chemical potential $\mu$ being the lowest eigenvalue of the
 Hartree-Fock Hamiltonian:
 \begin{align}
  \bigl(H_{HF}\bigr)_{ab}=&\bigl(H_{\Box}\bigr)_{ab}+2\bigl(H_{\rm int}^{2}
  \bigr)_{a_1a}^{a^\prime_1b}R_{a^\prime_1}R_{a_1}+ \nonumber \\
  &+4\bigl(H^{4h}_{\rm int}+H^{4v}_{\rm int}\bigr)_{a_1a_2a_3a}^{a^\prime_1
  a^\prime_2a^\prime_3b}\prod_{\mu=1}^3R_{a^\prime_\mu}R_{a_\mu}.
  \label{hf_hamiltonian}
 \end{align}
\end{subequations}
Once the amplitude $R_a$ is determined, the ground state energy (GSE) can be
computed using Eq. (\ref{energy_functional}).

Although we have formulated the HMF method for spin-$1/2$ systems, it can be
straightforwardly extended to higher spins as well.

Besides the GSE we will also be interested in computing the staggered
magnetization $M_z$, and the two-component VBS ``order parameter''
\cite{Sachdev_2008}:
\begin{align}
 {\rm Re}\Psi=&\frac{1}{N}\sum_{\bm x}(-1)^x\bS_{{\bm x}+{\bm e}_x}\bS_{\bm x};
 \nonumber \\
 {\rm Im}\Psi=&\frac{1}{N}\sum_{\bm x}(-1)^y\bS_{{\bm x}+{\bm e}_y}\bS_{\bm x},
 \nonumber
\end{align}
which allows us to characterize lattice point group symmetries of a state.

In the rest of this section we will sketch the HMF calculation of $E_0$, $M_z$
and $\Psi$ for the case of plaquettes, and only present final expressions for
the $4\times4$ clusters. The interested reader is referred to Ref.
\onlinecite{Isaev_2009}, where the technique is analyzed and developed in
greater detail. For simplicity we shall put $J\equiv1$.

\begin{figure}[!t]
 \begin{center}
  \includegraphics[width=\columnwidth]{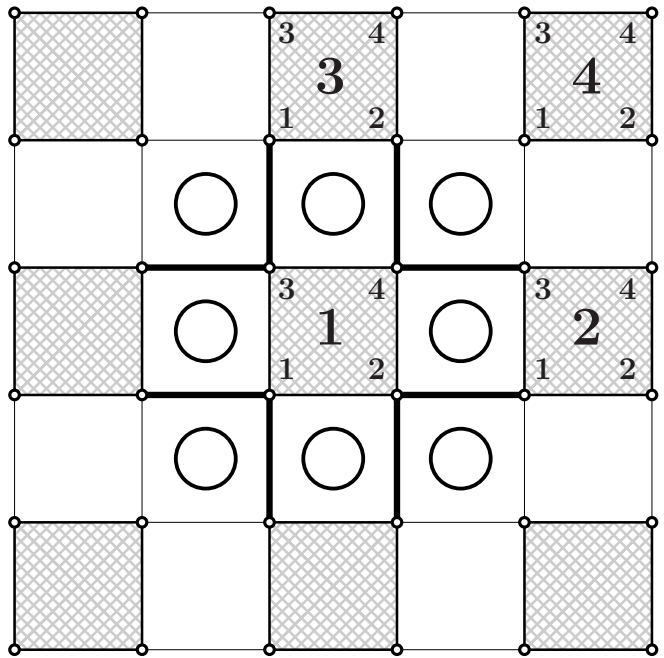}
 \end{center}
 \caption{The plaquette lattice. Thick lines denote interactions $J+Q/2$. The
          circles indicate four-spin terms of strength $Q$ in the Hamiltonian
          \eqref{initial_hamiltonian}. Small-sized numbers label spins within a
          plaquette, while the larger ones label plaquettes. }
 \label{2x2_lattice}
\end{figure}

\begin{figure}[!t]
 \begin{center}
  \includegraphics[width=\columnwidth]{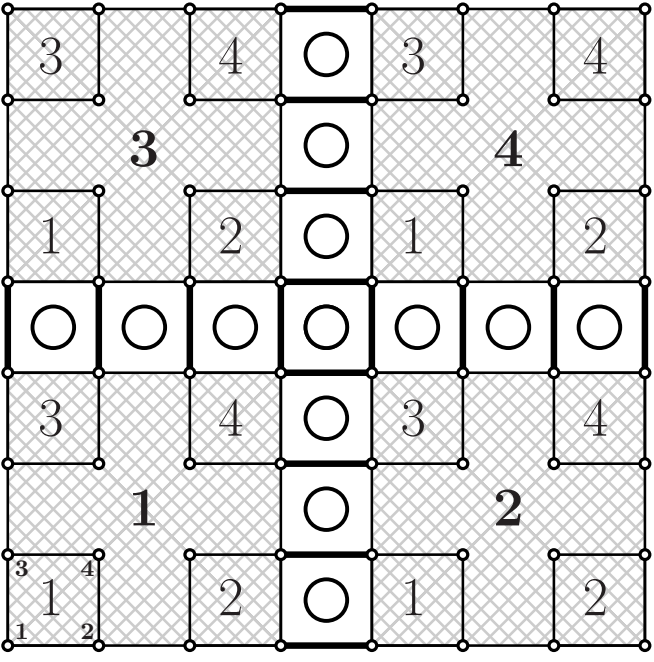}
 \end{center}
 \caption{Connectivity of the lattice formed by $4\times4$ spin clusters. Small
          circles indicate spins.Bold numbers label the $4\times4$ clusters,
          while thin ones denote plaquettes. Other notations, are the same as
          in Fig. \ref{2x2_lattice}.}
 \label{4x4_lattice}
\end{figure}

\subsection*{({\it i}) The plaquette degree of freedom}

We start by considering the simplest way to cover the lattice -- with
plaquettes, as shown in Fig. \ref{2x2_lattice}. At the same time we introduce
notations and concepts, which will be used in the following subsection.

The Hamiltonian for an isolated plaquette has the form:
\begin{align}
 H_\Box=&\biggl(1+\frac{Q}{2}\biggr)(\bS_1+\bS_4)(\bS_2+\bS_3)-
\nonumber \\
 &-Q\bigl[(\bS_1\bS_2)(\bS_3\bS_4)+(\bS_1\bS_3)(\bS_2\bS_4)\bigr].
 \label{bare_hamiltonian_2x2}
\end{align}
The interaction of this plaquette with the rest of the system can be
conveniently partitioned according to \eqref{initial_bosonic_hamiltonian} as:
\begin{displaymath}
 H_{\rm int}=H^2_{\rm int}+H^{4h}_{\rm int}+H^{4v}_{\rm int},
\end{displaymath}
where appropriately symmetrized individual terms are given by:
\begin{subequations}\label{interactions_2x2}
 \begin{align}
  H^2_{\rm int}=\frac{1+Q/2}{2}\bigl[&(\bS_{11}+\bS_{14})(\bS_{22}+\bS_{23})+
  \nonumber \\
  &+(\bS_{12}+\bS_{13})(\bS_{21}+\bS_{24})\bigr]- \label{2_interaction_2x2} \\
  -\frac{Q}{2}\bigl[
    (\bS_{11}&\bS_{12})(\bS_{23}\bS_{24})+(\bS_{11}\bS_{23})(\bS_{12}\bS_{24})+
  \nonumber \\
   +(\bS_{12}&\bS_{14})(\bS_{21}\bS_{23})+(\bS_{12}\bS_{21})(\bS_{14}\bS_{23})+
  \nonumber \\
   +(\bS_{13}&\bS_{14})(\bS_{21}\bS_{22})+(\bS_{13}\bS_{21})(\bS_{14}\bS_{22})+
  \nonumber \\
   +(\bS_{11}&\bS_{22})(\bS_{13}\bS_{24})+(\bS_{11}\bS_{13})(\bS_{22}\bS_{24})
  \bigr]; \nonumber
 \end{align}
 \begin{align}
  H^{4h}_{\rm int}=-\frac{Q}{8}&\bigl[(\bS_{14}\bS_{23}+\bS_{24}\bS_{13})
  (\bS_{32}\bS_{41}+\bS_{31}\bS_{42})+ \nonumber \\
  +(\bS_{34}&\bS_{43}+\bS_{44}\bS_{33})(\bS_{12}\bS_{21}+\bS_{11}\bS_{22})
  \bigr]; \label{4h_interaction_2x2}
 \end{align}
 \begin{align}
  H^{4v}_{\rm int}=-\frac{Q}{8}&\bigl[(\bS_{14}\bS_{32}+\bS_{12}\bS_{34})
  (\bS_{23}\bS_{41}+\bS_{21}\bS_{43})+ \nonumber \\
  +(\bS_{24}&\bS_{42}+\bS_{22}\bS_{44})(\bS_{13}\bS_{31}+\bS_{11}\bS_{33})
  \bigr]. \label{4v_interaction_2x2}
 \end{align}
\end{subequations}

It is convenient to work in the basis which diagonalizes the $Q$-independent
part of $H_\Box$, Eq. \eqref{bare_hamiltonian_2x2}. Such is, for instance, the
basis of eigenstates of the total angular momentum of the plaquette:
\begin{align}
 |&a\rangle=|l_1l_2LM\rangle, \label{basis_2x2} \\
 &\bl_1=\bS_1+\bS_4;\,\,\,\bl_2=\bS_2+\bS_3;\,\,\,{\bm L}=\bl_1+\bl_2.
 \nonumber
\end{align}
The matrix elements, which appear in Eq. \eqref{energy_functional}
\begin{align}
 \bigl(H^2_{\rm int}\bigr)^{a^\prime_1a^\prime_2}_{a_1a_2}\equiv&\langle
 a^\prime_1a^\prime_2|H^2_{\rm int}|a_1a_2\rangle; \nonumber \\
 \bigl(H^{4h,v}_{\rm int}\bigr)^{a^\prime_1a^\prime_2;a^\prime_3a^\prime_4}_
 {a_1a_2;a_3a_4}&\equiv
 \langle a^\prime_1a^\prime_2;a^\prime_3a^\prime_4|H^{4h,v}_{\rm int}|a_1a_2;
 a_3a_4\rangle \nonumber
\end{align}
can now be computed using the angular momentum addition theorems.

The staggered magnetization (along the $z$-axis) within a plaquette is given by
\begin{equation}
 M_z=\frac{1}{4}\bigl(S^z_1+S^z_4-S^z_2-S^z_3\bigr)_{ab}R_aR_b,
 \label{magnetization_2x2}
\end{equation}
while the function $\Psi$ can be written in the plaquette representation as:
\begin{align}
 {\rm Re}\Psi=&\frac{1}{N}\sum_i\bigl[\bS_{i1}\bS_{i2}+\bS_{i3}\bS_{i4}
 \bigr]- \nonumber \\
 &-\frac{1}{N}\sum_i\bigl[\bS_{i2}\bS_{i+\xh,1}+\bS_{i4}\bS_{i+\xh,3}
 \bigr]; \label{psi_2x2} \\
 {\rm Im}\Psi=&\frac{1}{N}\sum_i\bigl[\bS_{i1}\bS_{i3}+\bS_{i2}\bS_{i4}
 \bigr]- \nonumber \\
 &-\frac{1}{N}\sum_i\bigl[\bS_{i3}\bS_{i+\yh,1}+\bS_{i4}\bS_{i+\yh,2}
 \bigr]. \nonumber
\end{align}
In these equations the indices $i$ and $\hat{x}$ denote sites and basis vectors
of the plaquette lattice.

\begin{figure}[!t]
 \begin{center}
  \includegraphics[width=\columnwidth]{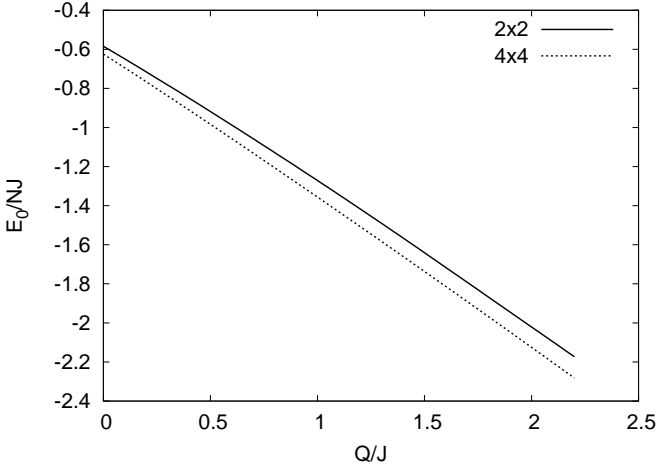}
 \end{center}
 \caption{Ground state energy as a function of $Q/J$ for plaquettes [case
          ({\it i})] and $4\times4$ spin clusters [case ({\it ii})].}
 \label{gse_2x2_4x4}
\end{figure}

\subsection*{({\it ii}) $4\times4$ spin clusters}

The coarse grained lattice obtained by choosing the $4\times4$ cluster as a
degree of freedom is shown in Fig. \ref{4x4_lattice}. Each spin operator
carries three indices: label of a cluster, label of a plaquette within this
cluster, and the position within this plaquette. Writing down the cluster
self-energy and the inter-cluster interactions is a straightforward, but
tedious task, which can be accomplished along the lines presented in the
previous subsection. Therefore, here we give only final expressions for $M_z$
and $\Psi$.

The staggered magnetization of a cluster is given by an equation analogous to
\eqref{magnetization_2x2}:
\begin{equation}
 M_z=\sum_{i=1}^4\bigl(S^z_{i1}+S^z_{i4}-S^z_{i2}-S^z_{i3}\bigr)_{A^\prime A}
 R_{A^\prime}R_A, \label{magnetization_4x4}
\end{equation}
where the summation extends over plaquettes within a cluster, and we used
capital indices $A$ to label states of the cluster. The function $\Psi$ can be
written as
\begin{align}
 {\rm Re}\Psi=\frac{1}{N}\sum_i&\bigl(\bS_{i11}\bS_{i12}+\bS_{i13}\bS_{i14}+
 \bS_{i31}\bS_{i32}+ \nonumber \\
 &+\bS_{i33}\bS_{i34}+\bS_{i21}\bS_{i22}+\bS_{i23}\bS_{i24}+ \nonumber \\
 &+\bS_{i41}\bS_{i42}+\bS_{i43}\bS_{i44}-\bS_{i12}\bS_{i21}- \nonumber \\
 &-\bS_{i14}\bS_{i23}-\bS_{i32}\bS_{i41}-\bS_{i34}\bS_{i43}\bigr)- \nonumber \\
 -\frac{1}{N}\sum_i&\bigl[\bS_{i22}\bS_{i+\xh,11}+\bS_{i24}\bS_{i+\xh,13}+
 \nonumber \\
 &+\bS_{i42}\bS_{i+\xh,31}+\bS_{i44}\bS_{i+\xh,33}\bigr]; \label{psi_4x4} \\
 {\rm Im}\Psi=\frac{1}{N}\sum_i&\bigl(\bS_{i11}\bS_{i13}+\bS_{i12}\bS_{i14}+
 \bS_{i21}\bS_{i23}+ \nonumber \\
 &+\bS_{i22}\bS_{i24}+\bS_{i31}\bS_{i33}+\bS_{i32}\bS_{i34}+ \nonumber \\
 &+\bS_{i41}\bS_{i43}+\bS_{i42}\bS_{i44}-\bS_{i13}\bS_{i31}- \nonumber \\
 &-\bS_{i14}\bS_{i32}-\bS_{i23}\bS_{i41}-\bS_{i24}\bS_{i42}\bigr)- \nonumber \\
 -\frac{1}{N}\sum_i&\bigl[\bS_{i33}\bS_{i+\yh,11}+\bS_{i34}\bS_{i+\yh,12}+
 \nonumber \\
 &+\bS_{i43}\bS_{i+\yh,21}+\bS_{i44}\bS_{i+\yh,22}\bigr]. \nonumber
\end{align}

\section{Results}
\label{section3}

\begin{figure}[!t]
 \begin{center}
  \includegraphics[width=\columnwidth]{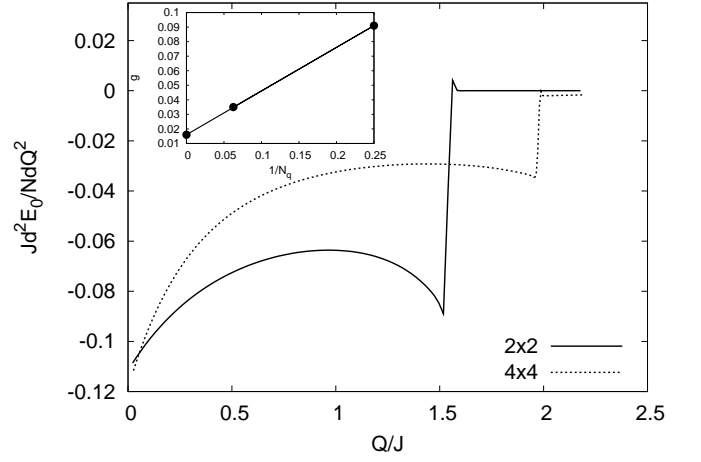}
 \end{center}
 \caption{Second-order derivative $d^2E_0/NdQ^2$ (main panel) as a function of
          $Q/J$ for cases ({\it i}) and ({\it ii}). The discontinuity at
          $Q/J\sim1.61$ ($2\times2$) and $Q/J\sim2.0$ ($4\times4$) indicates a
          second-order phase transition point. The inset shows the
          extrapolation of the jump $g=Jd^2E_0/NdQ^2|_{Q_c-0}^{Q_c+0}$ to
          $N_q\to\infty$.}
 \label{gse_d2_2x2_4x4}
\end{figure}

\begin{figure}[!b]
 \begin{center}
  \includegraphics[width=\columnwidth]{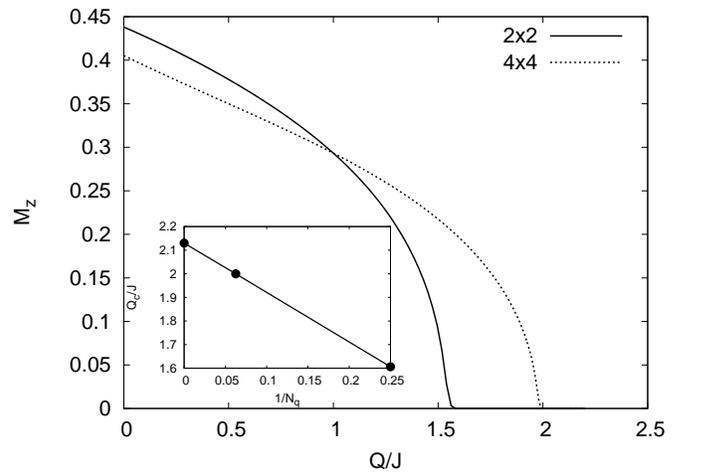}
 \end{center}
 \caption{Staggered magnetization (main panel), Eqs. \eqref{magnetization_2x2}
          and \eqref{magnetization_4x4}, as a function of $Q/J$ for cases
          ({\it i}) and ({\it ii}). The values of $Q_c$ are: $Q_c/J=1.61$ for
          case ({\it i}) and $Q_c/J=2.00$ for case ({\it ii}). The inset shows
          the scaling of $Q_c$.}
 \label{mag_2x2_4x4}
\end{figure}

We can now proceed with solution of the mean-field equation
\eqref{mean_field_eq}, supplemented by Eqs. \eqref{bare_hamiltonian_2x2},
\eqref{interactions_2x2} for the case ({\it i}) and analogous expressions in
the case ({\it ii}).

The physical quantities that we want to compute in the first place are the GSE
and the staggered magnetization. These are given by Eq.
\eqref{energy_functional} and Eqs. \eqref{magnetization_2x2} for the case ({\it
i}), and \eqref{magnetization_4x4} for the case ({\it ii}). In Fig.
\ref{gse_2x2_4x4} we present GSEs for both degrees of freedom. All energies
monotonically decrease with increasing $Q/J$ as a consequence of the negative
sign in front of the last term in Eq. \eqref{initial_hamiltonian}. At some
critical value of $Q=Q_c$ the system undergoes a phase transition from the
N\'eel state at small $Q$ to a spin-disordered state at $Q>Q_c$. This
transition can be seen either from the second derivative of the GSE,
$d^2E_0/NdQ^2$, shown in Fig. \ref{gse_d2_2x2_4x4}, or from the staggered
magnetization as a function of $Q/J$, presented in Fig. \ref{mag_2x2_4x4}.
Using these plots one obtains the numerical values $Q_c/J=1.61$ for plaquettes
and $Q_c/J=2.00$ for $4\times4$ clusters. Although the jump
$g=Jd^2E_0/NdQ^2|_{Q_c-0}^{Q_c+0}$ is numerically small, it remains finite:
$g\to0.016$, if extrapolated to the thermodynamic limit, based on these two
points (see the inset to Fig. \ref{gse_d2_2x2_4x4}). The finite-size scaling of
the critical point itself, presented in the inset to Fig. \ref{mag_2x2_4x4},
shows that $\lim_{N_q\to\infty}Q_c/J=2.13$. In order to demonstrate that our
results are reliable, we compute limiting values of the GSE and the
magnetization at $Q=0$: $\lim_{N_q\to\infty}E_0/NJ=-0.64$ and
$\lim_{N_q\to\infty}M_z=0.39$. These numbers should be compared to the accepted
QMC results \cite{Ceperley_1989}: $E_0/NJ=-0.67$ and $M_z=0.31$. We note,
finally, that due to few data points, the finite-size scalings presented here
are qualitative, and are intended to provide only an estimate for the
extrapolated quantities in the thermodynamic limit.

\begin{figure}[!t]
 \begin{center}
  \includegraphics[width=\columnwidth]{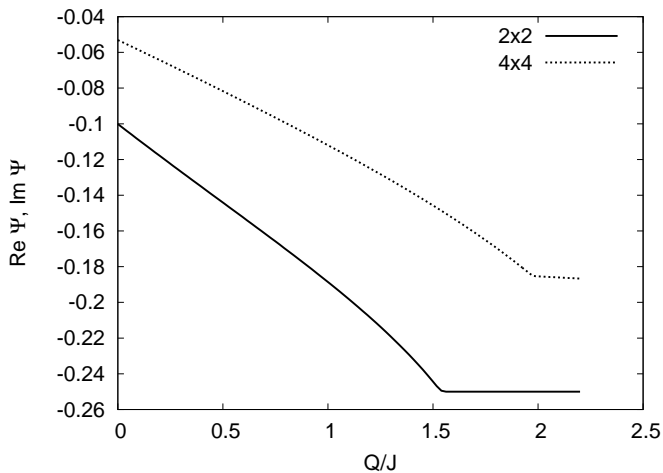}
 \end{center}
 \caption{The ``order parameter'' $\Psi$ for the two cases, studied in this
          paper. Notice the coincidence of curves for ${\rm Re}\Psi$ and
	  ${\rm Im}\Psi$. For $Q\geqslant Q_c$ this implies the plaquette
          nature of the quantum paramagnetic state.}
 \label{psi_2x2_4x4}
\end{figure}

Let us now discuss the symmetries of the various phases. The antiferromagnetic
state, which occurs for $Q<Q_c$, is known to preserve the lattice rotational
symmetry $C_4$, and spontaneously breaks the spin $SU(2)$ symmetry. The nature
of the paramagnetic phase, stabilized for $Q\geqslant Q_c$, can be unveiled by
computing expectation values of the function $\Psi$ given by Eqs.
\eqref{psi_2x2} and \eqref{psi_4x4} for cases ({\it i}) and ({\it ii}),
respectively. Although $\Psi$ is an integral quantity, it is sufficient for the
purpose of discriminating between plaquettized and dimerized ground states.
Namely, a plaquette phase preserves the four-fold lattice rotational symmetry,
implying
\begin{equation}\label{equality}
 {\rm Re}\Psi={\rm Im}\Psi,
\end{equation}
while in a dimerized state this equality does not hold. In Fig.
\ref{psi_2x2_4x4} we show ${\rm Re}\Psi$ and ${\rm Im}\Psi$. The equality
\eqref{equality} is satisfied throughout the phase diagram. This fact is not
surprising in the antiferromagnetic phase, but in the paramagnetic region it
presents a strong evidence against any type of dimerized ground states.
Although such states were allowed in the process of minimization, the
$C_4$-symmetric states always had lower energy. In fact, the ground state in
the non-magnetic region is a plaquette paramagnet, with each plaquette being in
its singlet ground state. However, due to the tensor nature of interactions in
\eqref{initial_hamiltonian} these plaquettes are interacting.

As already mentioned in the Introduction, the coarse graining procedure
explicitly breaks the lattice translation invariance, which should be
restored in the thermodynamic limit. Extrapolation to $N_q\to\infty$ shows that
${\rm Re}\Psi,{\rm Im}\Psi\to-0.04$, suggesting that the translation invariance
is indeed being recovered.

\section{Discussion}
\label{section4}

\begin{figure}[!t]
 \begin{center}
  \includegraphics[width=\columnwidth]{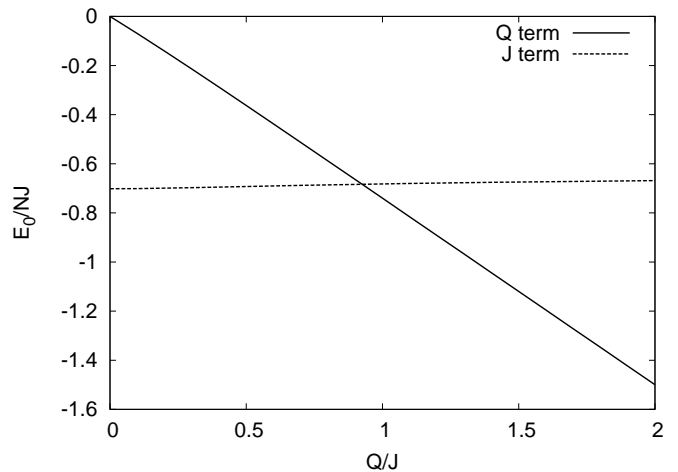}
 \end{center}
 \caption{Contributions to the GSE from the $J$ and $Q$ terms in
	  \eqref{initial_hamiltonian} for a $4\times4$ spin cluster with
          periodic boundary conditions. The unimportant term $-NQ/8$ is
          omitted.}
 \label{ed_4x4}
\end{figure}

Our calculations, presented in the previous section, demonstrate that the
Hamiltonian \eqref{initial_hamiltonian} exhibits a phase transition point
separating the N\'eel ordered state from a paramagnetic phase with broken
translational invariance, in agreement with conclusions of previous works
\cite{Sandvik_2007,Melko_2008_prl,Melko_2008_prb}. Most importantly, besides
establishing the existence of a phase transition, we were also able to unveil
the nature of the paramagnetic phase and show that a {\it correlated plaquette
state} is favored over a columnar dimer state which, although not conclusively,
seems to be preferred according to previous calculations \cite{Sandvik_2007}.

However, despite qualitative agreement, there is a quantitative discrepancy in
the numerical value of $Q_c$. Namely, the value obtained in the present paper
is much smaller than the one presented in Ref. \onlinecite{Sandvik_2007}.
Although we cannot provide a rigorous explanation for this discrepancy, we
would like to make some qualitative remarks in the following:

First of all, it is clear that the variational wavefunction
\eqref{HF_wavefunction}, being a low-density ansatz, generally leads to an
under-estimation of the four-boson scattering terms in Eq.
\eqref{initial_bosonic_hamiltonian}. In order to understand how significant
this error is and check that the results, presented in Figs.
\ref{gse_2x2_4x4}--\ref{mag_2x2_4x4}, are reasonable, we used data from the
exact diagonalization of $4\times4$ spin clusters to compare magnitudes of the
two terms: the ones, proportional to $J$ and $Q$, in Eq.
\eqref{initial_hamiltonian}. On physical grounds one would expect a phase
transition to occur when these terms become comparable. Figure \ref{ed_4x4}
presents the two contributions and their dependence on $Q/J$. Of course, the
crossing point at $Q/J\sim1$ does not determine the critical value $Q_c$, but
it provides a clue on where the phase transition may occur. Since the system is
gapped in the paramagnetic phase, one can argue that the size $4\times4$ is
large enough to describe the thermodynamic limit. Indeed, QMC data for $Q/J=10$
indicates that the GSE converges very rapidly with increasing system size
\cite{Sandvik_email}. Also, calculations analogous to that shown in Fig.
\ref{ed_4x4}, performed \cite{Sandvik_email} for systems up to $16\times16$
sites, indicate that the magnitude of the crossing point stays of order unity.  

Second, we would like to emphasize that, although there is no question
about the correctness of the QMC studies of Refs.
\onlinecite{Sandvik_2007,Melko_2008_prl,Melko_2008_prb,Chandrasekharan_2008}, 
the procedure used to extract physical quantities, like $Q_c$,
from the raw statistical data, is not straighforward and requires certain
assumptions \cite{Sandvik_2007}. Therefore, it is desirable to have
another independent determination of the phase transition point, for example,
from the data on staggered magnetization, computed in the entire $Q$ range.
While such calculations would definitely help to resolve this issue,
surprisingly, they have never been performed. QMC computations of finite
lattices does not suffer from the infamous sign problem in this case, thus it
yield better energy and magnetization values than the ones obtained here. Our
approach, on the other hand, focuses on establishing symmetry properties of
different phases, rather than improving numerical values for observable
quantities. It is this fact, which enables us to detect phase transition points
within a simple framework.

Our conclusions raise another important question regarding the nature of the
phase transition. We find it to be of the Landau type. Although the finite-size
scaling of the second-order derivative of the GSE, presented in the previous
section, displays a finite jump as $N_q\to\infty$, there is no way to
rigorously prove it. Thus, the possibility of a weakly first order transition
at $Q_c$ cannot be completely excluded. Indeed, in Ref.
\onlinecite{Chandrasekharan_2008} it was argued that this phase transition,
which was claimed to occur at the same point as in Ref.
\onlinecite{Sandvik_2007}, is of the first order. As any real-space method our
approach explicitly breaks translational invariance, and although the
finite-size scaling for $\Psi$ implies that this property is restored with
increasing cluster size, we cannot provide a rigorous symmetry-based analysis.

In summary, we determined the phase diagram of the $J$-$Q$ model
\eqref{initial_hamiltonian}, by using the recently proposed hierarchical
mean-field approach \cite{Isaev_2009,Ortiz_2003}. It was shown that there
exists a {\it single} (i.e. universal) mean-field framework (variational ansatz
for the ground state), which gives the complete phase diagram of the model. In
particular, we found that there exists a critical point at $Q_c\approx2.13J$,
which separates the antiferromagnetic phase from the non-magnetic state. The
latter breaks lattice translational invariance and was shown to represent a
correlated {\it plaquette paramagnetic} phase. Our results suggest that the
phase transition at $Q_c$ is of a Landau second order type, even in the
thermodynamic limit $N_q\to\infty$, although we cannot rigorously exclude the
possibility for it to become weakly first order.

We are indebted to A. W. Sandvik for bringing this problem to our attention and
numerous discussions and private communications in the course of completion of
this work. JD acknowledges support from the Spanish DGI under the grant
FIS2006-12783-C03-01.

\end{document}